\documentstyle[12pt,epsf]{article}

\catcode`\@=11
\long\def\@makefntext#1{
\protect\noindent \hbox to 3.2pt {\hskip-.9pt
$^{{\ninerm\@thefnmark}}$\hfil}#1\hfill}		

 \def\@makefnmark{\hbox to 0pt{$^{\@thefnmark}$\hss}}  

\def\ps@myheadings{\let\@mkboth\@gobbletwo
\def\@oddhead{\hbox{}
\rightmark\hfil\ninerm\thepage}
\def\@oddfoot{}\def\@evenhead{\ninerm\thepage\hfil
\leftmark\hbox{}}\def\@evenfoot{}
\def\sectionmark##1{}\def\subsectionmark##1{}}

\newcounter{sectionc}\newcounter{subsectionc}\newcounter{subsubsectionc}
\renewcommand{\section}[1] {\vspace{0.6cm}\addtocounter{sectionc}{1}
\setcounter{subsectionc}{0}\setcounter{subsubsectionc}{0}\noindent
	{\bf\thesectionc. #1}\par\vspace{0.4cm}}
\renewcommand{\subsection}[1] {\vspace{0.6cm}\addtocounter{subsectionc}{1}
	\setcounter{subsubsectionc}{0}\noindent
	{\it\thesectionc.\thesubsectionc. #1}\par\vspace{0.4cm}}
\renewcommand{\subsubsection}[1]{\vspace{0.6cm}\addtocounter{subsubsectionc}{1}
	\noindent {\rm\thesectionc.\thesubsectionc.\thesubsubsectionc.
	#1}\par\vspace{0.4cm}}

\newcounter{appendixc}
\newcounter{subappendixc}[appendixc]
\newcounter{subsubappendixc}[subappendixc]

\renewcommand{\appendix}[1] {\vspace{0.6cm}
        \refstepcounter{appendixc}
        \setcounter{figure}{0}
        \setcounter{table}{0}
        \setcounter{equation}{0}
        \renewcommand{\thefigure}{\Alph{appendixc}.\arabic{figure}}
        \renewcommand{\thetable}{\Alph{appendixc}.\arabic{table}}
        \renewcommand{\theappendixc}{\Alph{appendixc}}
        \renewcommand{\theequation}{\Alph{appendixc}.\arabic{equation}}
        \noindent{\bf Appendix \theappendixc #1}\par\vspace{0.4cm}}

\def\abstracts#1{{
	\centering{\begin{minipage}{30pc}\tenrm\baselineskip=12pt\noindent
	\centerline{\tenrm ABSTRACT}\vspace{0.3cm}
	\parindent=0pt #1
	\end{minipage}}\par}}

\renewenvironment{thebibliography}[1]
	{\begin{list}{\arabic{enumi}.}
	{\usecounter{enumi}\setlength{\parsep}{0pt}
\setlength{\leftmargin 1.25cm}{\rightmargin 0pt}
	 \setlength{\itemsep}{0pt} \settowidth
	{\labelwidth}{#1.}\sloppy}}{\end{list}}
\newcounter{itemlistc}
\newcounter{romanlistc}
\newcounter{alphlistc}
\newcounter{arabiclistc}

\newcommand{\fcaption}[1]{
        \refstepcounter{figure}
        \setbox\@tempboxa = \hbox{\tenrm Fig.~\thefigure. #1}
        \ifdim \wd\@tempboxa > 6in
           {\begin{center}
        \parbox{6in}{\tenrm\baselineskip=12pt Fig.~\thefigure. #1}
            \end{center}}
        \else
             {\begin{center}
             {\tenrm Fig.~\thefigure. #1}
              \end{center}}
        \fi}
\newcommand{\tcaption}[1]{
        \refstepcounter{table}
        \setbox\@tempboxa = \hbox{\tenrm Table~\thetable. #1}
        \ifdim \wd\@tempboxa > 6in
           {\begin{center}
        \parbox{6in}{\tenrm\baselineskip=12pt Table~\thetable. #1}
            \end{center}}
        \else
             {\begin{center}
             {\tenrm Table~\thetable. #1}
              \end{center}}
        \fi}

\def\@citex[#1]#2{\if@filesw\immediate\write\@auxout
	{\string\citation{#2}}\fi
\def\@citea{}\@cite{\@for\@citeb:=#2\do
	{\@citea\def\@citea{,}\@ifundefined
	{b@\@citeb}{{\bf ?}\@warning
	{Citation `\@citeb' on page \thepage \space undefined}}
	{\csname b@\@citeb\endcsname}}}{#1}}
\newif\if@cghi
\def\cite{\@cghitrue\@ifnextchar [{\@tempswatrue
	\@citex}{\@tempswafalse\@citex[]}}
\def\citelow{\@cghifalse\@ifnextchar [{\@tempswatrue
	\@citex}{\@tempswafalse\@citex[]}}
\def\@cite#1#2{{$\null^{#1}$\if@tempswa\typeout
	{IJCGA warning: optional citation argument
	ignored: `#2'} \fi}}

\def\fnt#1#2{\footnotetext{\kern-.3em
	{$^{\mbox{\sevenrm #1}}$}{#2}}}
 1
 1
 1

\font\elevenbf=cmbx10     scaled\magstephalf

\font\elevenit=cmti10     scaled\magstephalf

\font\tenbf=cmbx10
\font\tenrm=cmr10
\font\tenit=cmti10

\font\ninerm=cmr9

\renewcommand{\Re}{{\rm Re}}
\renewcommand{\Im}{{\rm Im}}

\newcommand{\BLDgamma}{\gamma\hspace{-1.24ex}\gamma
                            \hspace{-1.24ex}\gamma}
\newcommand{\bfp}{{\bf p}}
\newcommand{\rmp}{{\rm p}}
\newcommand{\bfk}{{\bf k}}
\newcommand{\rmk}{{\rm k}}

\newcommand{\bfx}{{\bf x}}
\newcommand{\Z}{{\rm Z}}
\newcommand{\calc}{{\cal C}}
\newcommand{\slsh}[1]{\mbox{$#1 \hspace{-.45em} /$}}
\def\bld#1{\mbox{\boldmath$#1$}}
\def\bldsml#1{\mbox{\boldmath\scriptsize$#1$}}
\begin{document}
%
\mbox{}
\vspace{-7.em}
\begin{flushright}
\bf
TTP95-47\footnotemark[3]\\
December 1995\\
hep-ph/9512409
\end{flushright}
\vspace{1.em}
\begin{center}
{\tenbf  POLARIZED TOP QUARKS\footnotemark[1] \footnotemark[2]}\\
\vspace{0.8cm}
\baselineskip=14pt
{\tenrm M. JE\.ZABEK\\}
{\tenit Institute of Nuclear Physics, Kawiory 26a, PL-30055 Cracow,
Poland\\}
\vglue 0.2cm
{\tenrm and\\}
\vglue 0.2cm
{\tenrm R. HARLANDER, J. H. K\"UHN and M. PETER\\}
{\tenit Institut f\"ur Theoretische Teilchenphysik, Universit\"at
Karlsruhe\\}
{\tenit D-76128 Karlsruhe, Germany\\}
\end{center}
\footnotetext[1]{
\ninerm
\baselineskip=11pt
Work partly supported by Polish State Committee for Scientific
Research (KBN) grants 2P3025206 and 2P30207607.}
\footnotetext[2]{
Invited talk presented at the {\it Workshop on Physics and Experiments with
Linear Colliders}, September 8--12, 1995, Morioka--Appi, Iwate, Japan,
	to appear in the proceedings.}
\footnotetext[3]{The complete paper is also available via anonymous ftp at
ftp://www-ttp.physik.uni-karlsruhe.de/, or via www at
http://www-ttp.physik.uni-karlsruhe.de/cgi-bin/preprints/.}
\baselineskip=12pt
\vspace{0.9cm}
\abstracts{
\noindent
Recent calculations are presented of top quark polarization in
$t\bar t$ pair production close to threshold. S--P-wave interference
gives contributions to all components of the top quark
polarization vector. Rescattering of the decay products is considered.
Moments of the fourmomentum of the charged lepton in semileptonic
top decays are calculated and shown to be very sensitive to the top
quark polarization.}

\vspace{-3.ex}
\vfil
\rm\baselineskip=14pt
\section{Introduction}
Threshold production of top quarks at a future
electron--positron
collider will allow to study their properties with
extremely
high precision. The dynamics of the top quark
is strongly influenced by its large width
$\Gamma_t\approx 1.5$ GeV.  Individual quarkonium resonances
can no longer be resolved, hadronization effects
are irrelevant and an effective cutoff of the
large distance (small momentum) part of the hadronic interaction
is introduced\cite{K,BDKKZ,FadKhoz1}.
This in turn allows to measure the short distance part
of the potential, leading to a precise determination
of the strong coupling constant\cite{strassler}.
The analysis of the total cross section combined with
its momentum distribution
will determine its mass with an accuracy of at least 300 MeV and its
width to
about 10\%. For a Higgs boson mass of order 100 GeV
even the $t\bar t H$ Yukawa coupling could be indirectly
deduced from its contribution to the vertex
correction\cite{work}.
Additional constraints on these parameters can be derived
from the forward--backward asymmetry of top quarks and from
measurements of the top quark spin.
Close to threshold, for $E=\sqrt{s}-2m_t\ll m_t$,
the total cross section and similarly the momentum
distribution of the quarks are essentially governed by
the $S$-wave amplitude, with $P$-waves suppressed
$\sim \beta^2 \sim
\sqrt{E^2+\Gamma_t^2}/m_t \approx 10^{-2}$.
The forward--backward asymmetry and, likewise,
the transverse component of the top quark spin originate
from the interference between $S$- and $P$-wave
amplitudes and are, therefore, of order $\beta \approx 10^{-1}$
even close to threshold.  Note that the expectation value
of the momentum is always different
from zero as a consequence of the large top width
and the uncertainty principle, even for $E=0$.
\par\noindent
It has been demonstrated\cite{FadKhoz1,strassler}
that the Green function technique is particularly
suited to calculate the total cross section in the threshold region.
The method has been extended\cite{JKT,MurSum1}
to predict the top quark momentum distribution.
A further generalization then leads to the inclusion of
$P$-waves and, as a consequence, allows to predict
the forward--backward asymmetry\cite{MurSum2}.
It has been shown\cite{hjkt}  that the same function
$\varphi_{_R}(\bfp,E)$ which results from the $S$--$P$-wave
interference governs the dynamical behaviour of the forward--backward
asymmetry as well as the angular dependence of the transverse
part of the top quark polarization.  The close
relation between this result and the tree level prediction,
expanded up to linear terms in $\beta$, has been emphasised.
The relative importances of $Z$ versus $\gamma$ and
of axial versus vector couplings depend on the
electron (and/or positron) beam polarization.
All predictions can, therefore, be further tested by exploiting
their dependence on beam polarization.
In fact the reaction $e^+e^-\to t\bar t$ with longitudinally
polarized beams is the most efficient and flexible source
of polarized top quarks. At the same time the longitudinal
polarization of the electron beam is an obvious option for
a future linear collider.
Recently\cite{HJKP} these results have been expanded in two directions:
\begin{itemize}
\item {\it Normal polarization}.
Calculation of the polarization normal to the production plane is a
straightforward extension of the previous work\cite{hjkt}
and is based
on the same nonrelativistic Green function as before,
involving, however, the imaginary part of the interference term
$\varphi_{_I}(\bfp,E)$.
A component of the top quark polarization normal to the production
plane may also be induced by time reversal odd components of
the $\gamma t \bar t$- or $\Z t \bar t$-coupling with an electric
dipole moment as most prominent example. Such an effect would be
a clear signal for physics beyond the standard model.
The relative sign of particle versus antiparticle polarizations
is opposite for the QCD-induced and the $T$-odd terms respectively,
which allows to discriminate the two effects. Nevertheless it is
clear that a complete understanding of the QCD-induced component is
mandatory for a convincing analysis of the $T$-odd contribution.
\item {\it Rescattering}.
Both $t$ quark and $\bar t$ antiquark are unstable and decay
into $W^+b$ and $W^-\bar b$, respectively.
Neither $b$ nor $\bar b$ can be
considered as freely propagating particles.
Rescattering in the $t\bar b$ and $b\bar t$ systems
affects not only the momenta of the decay
products but also the polarization of the top quark.
Moreover, in the latter case, when the top quark decays first and its
colored decay product $b$ is rescattered in a Coulomb-like chromostatic
potential of the spectator $\bar t$, the top polarization is not
a well defined quantity. Instead one can consider other
quantities, like the total angular momentum of
the $Wb$ subsystem, which are equal to the spin of top quark
in the situation when rescattering is absent.
These rescattering corrections are suppressed by $\alpha_s$.
The resulting modifications of the momentum distribution
are therefore relatively minor and as far
as the total cross section is concerned can even
be shown to vanish\cite{MY94}. In contrast the forward--backward
asymmetry as well as the transverse and normal parts of the
top quark spin are suppressed by a factor $\sim \beta$.
Thus, they are relatively more sensitive
towards rescattering corrections.
Rescattering in the $b\bar b$ system is less important and will be
neglected.
\end{itemize}
\par\noindent
It is well known\cite{teupitz} that the direction of the charged
lepton in semileptonic decays is the best polarization analyzer
for the top quark. The reason is\cite{JK89b} that in the top quark
rest frame the double differential energy--angular distribution
of the charged lepton is a product of the energy and the angular
dependent factors. The angular dependence is of the form
$(1 + P\cos\theta)$, where $P$ denotes the top quark polarization
and $\theta$ is the angle between the polarization three-vector
and the direction of the charged lepton. Gluon radiation and
virtual corrections in  the top quark decay practically do not
affect these welcome properties\cite{CJK91}. It is therefore
quite natural to perform polarization studies by measuring
the inclusive distributions of say $\mu^+$
in the process $e^+e^-\to t(\mu^+\nu_\mu b)\bar t(jets)$.
This can be also convenient from the experimental point of
view because there is no missing energy-momentum for the $\bar t$
subsystem. From the theoretical point of view the direction of the
charged lepton can be considered as another quantity which is
equivalent to the top quark polarization when rescattering is
absent. Of course, it is well defined also
in the case of $b\bar t$ rescattering.
However, the semi-analytic calculation of the latter contribution
is a very difficult task because production and decay mechanisms
are coupled. A way out\cite{HJKP} is to calculate moments of the charged
lepton four-momentum distributions. The results of this analysis
are published elsewhere\cite{HJKP}.

\section{Green functions, angular distributions
        and quark polarization\label{technic_sec}}
\subsection{The nonrelativistic limit\label{nonrel_subs}}
Top quark production in the threshold region is conveniently
described by the Green function method which allows
to introduce in a natural way the effects of the large top
decay rate $\Gamma_t$ and avoids the summation of many
overlapping resonances. The total cross section can be
obtained from the imaginary part of the Green function
$G(\bfx=0,\bfx'=0,E)$ via the optical theorem. To predict
the differential momentum distribution, however, the
complete $\bfx$-dependence of $G(\bfx,\bfx'=0,E)$ (or, more precisely,
its Fourier transformed) is required. In a calculation with
non-interacting quarks close to threshold the
forward--backward asymmetry, the leading angular dependent term
$\sim \cos\vartheta$ and the transverse part of the top quark
polarization are all proportional to the quark velocity
$\beta$ and originate from the interference of a $S$-wave
with a $P$-wave amplitude.
These distributions are described
by $\nabla' \cdot G(\bfx,\bfx',E)|_{\bfx'=0}$ or,
equivalently, by the component of the Green
function with angular momentum one. The connection between
the relativistic treatment and the nonrelativistic
Lippmann--Schwinger equation has been discussed in
the literature\cite{strassler,MurSum2}. The subsequent discussion
follows these lines.  It includes, however, also the spin degrees
of freedom and is, furthermore, formulated sufficiently general
such that it is immediately applicable to other reactions of interest.
The main ingredient in the derivation of the nonrelativistic limit
is the ladder approximation for the vertex function $\Gamma_\calc$.
This vertex function is the solution of the following integral equation
depicted in Fig.\ref{lipp.ps}:
\begin{figure}
\begin{center}
  \leavevmode
  \epsfxsize=12.cm
  \epsffile[50 220 540 350]{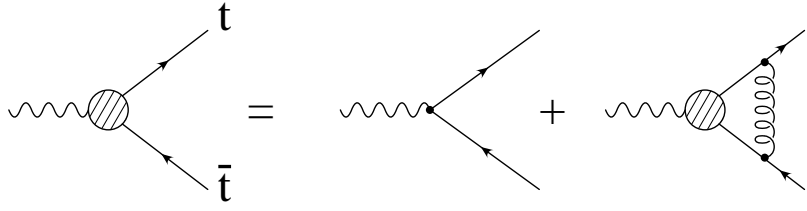}\\
  \caption[]{
	\label{lipp.ps}\sloppy Lippmann--Schwinger equation in diagrammatical
     form.}
\end{center}
\end{figure}
\begin{equation}\label{int_eq}
  \Gamma_\calc = \calc + \int {d^4k \over (2 \pi)^4}
  \left( -{4 \over 3} 4 \pi \alpha_s \right)
    D_{\mu \nu}(p-k) \gamma^\mu S_{\rm F}(k+{q \over 2})
    \Gamma_\calc(k,q) S_{\rm F}(k-{q\over2}) \gamma^\nu ,
\end{equation}
%
\begin{figure}[htb]
  \begin{flushleft}
    \leavevmode
    \epsfxsize=5.cm
    \epsffile[165 200 400 370]{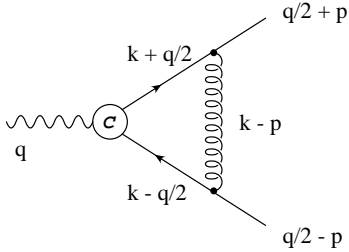}\\[-1.cm]
    \hfill
    \parbox{8.cm}{
      \caption[]{\label{1loopc.ps}\sloppy Definition of the four-momenta.}}
  \end{flushleft}
\end{figure}
with $\calc=\gamma_\mu$ or $\gamma_\mu\gamma_5$
in the cases of interest for $e^+e^-$-annihilation.
The conventions for the flow of momenta are illustrated in
Fig.\ref{1loopc.ps}.
The four-momenta are related to the nonrelativistic
variables by
\begin{eqnarray}
q &=& (2 m_t + E,{\bf 0}) \nonumber\\
p &=& (0, \bfp) \label{impulse}\\
k &=& (k_0,\bfk) .\nonumber
\end{eqnarray}
In perturbation theory the ladder approximation is motivated
by the observation that for each additional rung the energy
denominator after loop integration compensates the coupling
constant attached to the gluon propagator. This is demonstrated most
easily in Coulomb gauge. Contributions from transverse gluons
as well as those from other diagrams are suppressed by higher powers of
$\beta\sim\alpha_s$.
The gluon propagator is thus replaced by the
instantaneous nonrelativistic potential
\begin{equation}\label{gluon_prop}
{4\over 3}\, 4\pi \alpha_s D_{\mu\nu}(p) \rightarrow
         i\, V(\bfp)\, \delta_{\mu 0} \delta_{\nu 0} .
\end{equation}
The dominant contribution to the integral originates from the region
where $|\bfk|\ll m_t$. Including terms linear in $\bfk$,
quark and antiquark
propagators are approximated by\\
\parbox{32.em}{
\begin{eqnarray*}
\hspace{10.ex}
S_{\rm F}(k+{q\over2}) &=& i {\Lambda_+ -
     {\bfk\cdot\bldsml\gamma \over 2 m_t} \over
     {E \over 2} + k_0 - {\bfk^2 \over 2 m_t} + i {\Gamma_t \over 2}} \\
S_{\rm F}(k-{q\over2}) &=& i {\Lambda_- -
      {\bfk\cdot \bldsml\gamma \over 2 m_t} \over
      {E \over 2} - k_0 - {\bfk^2 \over 2 m_t} + i {\Gamma_t \over 2}} \, ,\\
\Lambda_\pm &=& {1\pm\gamma^0\over 2} \, .
\end{eqnarray*}}
\hfill\parbox{5.ex}{
        \begin{eqnarray} \label{fermion_props} \end{eqnarray}}
The ``elementary'' vertex $\calc$ is independent of $k_0$.  (Within the
present approximations this is even true if $\calc$ does depend on $\bfk$ as is
the case in the analogous treatment of $\gamma\gamma\to t\bar t$
discussed below.)
Up to and including order $\beta$ terms
a selfconsistent solution of the integral equation (\ref{int_eq})
can be obtained if $\Gamma_\calc$ is taken
independent of $k_0$ and the nonrelativistic spins of $t$ and $\bar t$.
The $k_0$ integration is then easily performed and the
integral equation simplified to
\begin{equation}\label{int_eq_2}
\Gamma_\calc = \calc + \int {d^3 k \over (2 \pi)^3}
        V(\bfp - \bfk) \gamma^0
        (\Lambda_+ - {\bfk\cdot \BLDgamma \over 2 m_t})
        {\Gamma_\calc(\bfk,E) \over E - {\bfk^2 \over m_t} + i \Gamma_t}
        (\Lambda_- - {\bfk\cdot \BLDgamma \over 2 m_t})
        \gamma^0  .
\end{equation}
In the calculation of the cross section for the production of
$t$ plus $\bar t$
with momenta $q/2\pm p$ and spins $s_\pm$ respectively traces ${\cal H}$ of the
following structure will arise:
\begin{eqnarray}
{\cal H} &=& {\rm Tr} \{ {\cal P}_+({q \over 2} + p,s_+)
  \Gamma_\calc {\cal P}_-({q \over 2} - p,s_-) \bar{\Gamma}_{\calc'} \} , \\
  {\cal P}_\pm(p,s) &=& {\slsh p \pm m_t \over 2 m_t}\,
        {1+\gamma_5 \slsh s \over 2} ,
\end{eqnarray}
where we allowed for mixed terms with $\calc$ different from $\calc'$.
Expanding again up to terms linear in $\bfk$, this trace can be transformed
into
\begin{eqnarray}
{\cal H} &=&
        {\rm Tr} \{ {\cal S}_+ \widetilde\Gamma_\calc {\cal S}_-
        \bar{\widetilde\Gamma}_{\calc'} \} , \\
{\cal S}_\pm &=& {1 \pm {\bf s}_\pm \cdot\bld{\Sigma} \over 2}  ,\\
\bld{\Sigma} &=& \BLDgamma \gamma_5 \gamma^0 =
        \left(\begin{array}{cc} \bld{\sigma} & 0 \\
                                0            & \bld{\sigma}
                \end{array}
        \right) ,
\end{eqnarray}
with the nonrelativistic reduction defined through
\begin{equation}\label{tilde}
\widetilde{\Gamma}_\calc(\bfp,E)
        = \Lambda_+ \, (1-{\bfp\cdot \BLDgamma \over 2 m_t}) \,
        \Gamma_\calc(\bfp,E)\,(1-{\bfp\cdot \BLDgamma \over 2 m_t}) \,
        \Lambda_-  .
\end{equation}
It is thus sufficient to calculate the ``reduced'' vertex
function $\widetilde \Gamma_\calc$. Dropping again terms
of order $\bfk^2$, the corresponding integral
equation is cast into a particularly simple form
\begin{equation}\label{int_eq_3}
\widetilde{\Gamma}_\calc(\bfp,E) = \widetilde{\calc} (\bfp) +
        \int {d^3 k \over (2 \pi)^3} V(\bfp - \bfk)
        {\widetilde{\Gamma}_\calc(\bfk,E) \over E -
        {\bfk^2 \over m_t} + i \Gamma_t}  .
\end{equation}
Consistent with the nonrelativistic approximation only the
constant and the linear term in the Taylor expansion of the
elementary vertex will be considered\footnote{
In the notation of K\"uhn et al.\cite{kks} one gets:
  $\widetilde\calc(0)=\Lambda_+{\cal O}_0\Lambda_-$ and
  ${\bf D} = \Lambda_+\left[-{1\over 2m_t}\{{\cal O}_0,\BLDgamma\}_+ +
	\hat{\bld{\cal O}}\right]\Lambda_-$.}
\begin{equation}
\widetilde{\calc}(\bfp) = \widetilde{\calc}(0) + {\bf D}\cdot \bfp
\end{equation}
The matrices  $\widetilde\calc(0)$ and ${\bf D}$ may in general
depend on external momenta, polarization vectors or Lorentz
indices. A selfconsistent solution for the vertex
$\widetilde\Gamma_\calc$ is then given by
\begin{equation}\label{ansatz}
\widetilde\Gamma_\calc(\bfp,E) = \widetilde\calc (0)
{\cal K}_{\rm S}(\rmp,E) +
        {\bf D}\cdot \bfp \, {\cal K}_{\rm P}(\rmp,E)
\end{equation}
The scalar vertex functions ${\cal K}_{\rm S,P}$ depend on
$$  \rmp = |\bfp| $$
and $E$ only.
They are solutions of the nonrelativistic integral equations
\begin{eqnarray}
{\cal K}_{\rm S} (\rmp,E) &=& 1 + \int {d^3 k \over (2 \pi)^3}
        V(\bfp - \bfk) {{\cal K}_{\rm S} (\rmk,E) \over E -
        {\bfk^2 \over m_t} + i \Gamma_t}  \\
{\cal K}_{\rm P} (\rmp,E) &=& 1 + \int {d^3 k \over (2 \pi)^3}
        {\bfp\cdot \bfk \over \bfp^2}
        V(\bfp - \bfk) {{\cal K}_{\rm P} (\rmk,E) \over E -
        {\bfk^2 \over m_t} + i \Gamma_t}
\end{eqnarray}
and are closely related to the Green function
${\cal G}(\bfp,\bfx,E)$ which, in turn, is a solution of the
Lippmann--Schwinger equation
\begin{equation}\label{LS}
\bigg[E - {\bfp^2 \over m_t} + i \Gamma_t \bigg]
        {\cal G} (\bfp,\bfx,E) = e^{i \bfp \cdot \bfx} +
        \int {d^3 k \over (2 \pi)^3} V(\bfp - \bfk)
        {\cal G} (\bfk,\bfx,E) .
\end{equation}
Let us denote the first two terms of the Taylor series with respect
to $\bfx$ by $G$ and $F$ respectively:
\[
{\cal G}(\bfp,\bfx,E) =
G(\rmp,E) + \bfx \cdot \bfp \, F(\rmp,E) + \ldots
\]
They are solutions of the integral equations
\begin{eqnarray}
G(\rmp,E) &=& G_0(\rmp,E) + G_0(\rmp,E) \int {d^3 k\over (2 \pi)^3}
        V(\bfp - \bfk) G(\rmk,E)                   \label{lipps} \\
F(\rmp,E) &=& G_0(\rmp,E) + G_0(\rmp,E) \int {d^3 k\over (2 \pi)^3}
        {\bfp \cdot \bfk \over \bfp^2}
        V(\bfp - \bfk) F(\rmk,E)  ,                \label{lippp}
\end{eqnarray}
with
\begin{equation}\label{free}
G_0(\rmp,E) = {1\over E - {\bfp^2 \over m_t} + i \Gamma_t} ,
\end{equation}
and the relation between Green function and vertex function
\begin{equation}
  G(\rmp,E) = G_0(\rmp,E) {\cal K}_{\rm S} (\rmp,E) , \qquad
  F(\rmp,E) = G_0(\rmp,E) {\cal K}_{\rm P} (\rmp,E)
\end{equation}
is  evident.
In the case of $e^+e^-$-annihilation top production proceeds
through the space components of the vector and axial vector
current.  The relevant elementary
vertex $\widetilde\calc(\bfp)$ is given by
\begin{eqnarray}
\widetilde{\gamma_j}(\bfp) &=& \Lambda_+ \gamma_j \Lambda_- \\
\widetilde{\gamma_j\gamma_5}(\bfp) &=& \Lambda_+({i\over m_t})
        (\bld{\gamma}\times\bfp)_j \Lambda_-
\end{eqnarray}
for vector and axial current respectively. Production of $t\bar t$ in
$\gamma\gamma$-fusion would lead to an elementary vertex of the form
\begin{eqnarray}
  \widetilde{\calc}(0) &\propto& i (\bld{\epsilon}_1
  \times \bld{\epsilon}_2)\,
  {\bf n}_{e^-} \Lambda_+ \gamma_5 \Lambda_-\\
  {\bf D} &\propto& {1\over m_t} \Lambda_+ \left[
  (\bld{\epsilon}_1 \cdot \bld{\epsilon}_2)
  ({\bf n}_{e^-}\cdot \BLDgamma) {\bf n}_{e^-} +
  (\bld{\epsilon}_2\cdot \BLDgamma) \, \bld{\epsilon}_1 +
  (\bld{\epsilon}_1\cdot \BLDgamma) \, \bld{\epsilon}_2 \right]
  \Lambda_-,
\end{eqnarray}
with $\bld{\epsilon}_1$, $\bld{\epsilon}_2$
the polarization vectors of the
photons,
and the present formalism applies equally well. This case has
been studied by Fadin et al.\cite{FKotsky}.
\par\noindent

\subsection{Top production in electron positron annihilation
\label{eplus_subs}}
With these ingredients it is straightforward to calculate the
differential momentum distribution and the polarization of
top quarks produced in electron positron annihilation.
Let us introduce the following conventions for the fermion
couplings
\begin{equation}
v_f = 2 I^3_f - 4 q_f \sin^2\theta_W , \qquad  a_f = 2 I^3_f .
\end{equation}
$P_\pm$ denotes the longitudinal electron/positron
polarization and
\begin{equation}\label{chi}
\chi={P_+-P_-\over1-P_+P_-}
\end{equation}
can be interpreted as effective longitudinal polarization of
the virtual intermediate photon or $Z$ boson.
The following abbreviations will be useful below:
\begin{eqnarray*}
a_1 &=& q_e^2 q_t^2 + (v_e^2 + a_e^2) v_t^2 d^2 +
	2 q_e q_t v_e v_t d \\
a_2 &=& 2 v_e a_e v_t^2 d^2 + 2 q_e q_t a_e v_t d \\
a_3 &=& 4 v_e a_e v_t a_t d^2 + 2 q_e q_t a_e a_t d \\
a_4 &=& 2 (v_e^2 + a_e^2) v_t a_t d^2 + 2 q_e q_t v_e a_t d \\
d &=& {1\over 16 \sin^2\theta_W\cos^2\theta_W}\,{s\over s - M_Z^2}.
\end{eqnarray*}
The differential cross section, summed over polarizations of quarks
and expanded up to terms linear in $\bfp$, is thus given by
\begin{eqnarray}
{d\sigma \over d\bfp} &=&
{3 \alpha^2 \Gamma_t \over 4 \pi m_t^4} (1-P_+P_-)
\left[ { (a_1 + \chi a_2)
	\left(1-{16 \alpha_s \over 3 \pi} \right)
	\left|G(\rmp,E)\right|^2 + }\right. \nonumber\\
& & 
\left. {+(a_3+\chi a_4)
\left( 1-{12\alpha_s \over 3 \pi} \right)
{\rmp \over m_t} \Re \left(\,G(\rmp,E) F^*(\rmp,E)\,\right)\,
\cos\vartheta} \right]  \label{dsig_d3p} .
\end{eqnarray}
The vertex corrections from hard gluon exchange for $S$-wave\cite{barbieri}
and $P$-wave\cite{KZ} amplitudes are included in this
formula. It leads to the following forward--backward asymmetry
\begin{equation}\label{afb}
{\cal A}_{\rm FB}(\rmp,E) = C_{\rm FB}(\chi)\, \varphi_{_R}(\rmp,E),
\end{equation}
with
\begin{equation}
   C_{\rm FB}(\chi) = {1 \over 2}\, {a_3
    + \chi a_4 \over a_1 + \chi a_2} ,
\end{equation}
$\varphi_{_R}(\rmp,E) = \Re\,\varphi$,  and
\begin{equation}\label{phi}
\varphi(\rmp,E) =
{(1-{4 \alpha_s/3 \pi})\over (1-{8 \alpha_s/3 \pi})}\,
	{\rmp \over m_t}\,
	{F^* \!(\rmp,E) \over G^* \!(\rmp,E)}  .
\end{equation}
This result is still differential in the top quark momentum.
Replacing $\varphi(\rmp,E)$ by
\begin{equation}\label{cap_phi}
\Phi(E) =
{(1-{4 \alpha_s/3 \pi})\over (1-{8 \alpha_s/3 \pi})}\,
{\int_0^{\rmp_m} d\rmp\,
{\rmp^3 \over m_t}\, F^*(\rmp,E)G(\rmp,E) \over
\int_0^{\rmp_m} d\rmp \, \rmp^2 \left|G(\rmp,E)\right|^2}  .
\end{equation}
one obtains the integrated forward--backward
asymmetry\footnote{For the case without beam polarization
this coincides with the earlier result\cite{MurSum2}, as far as the
Green function is concerned. It differs, however, in the
correction originating from hard gluon exchange.}.
The cutoff $\rmp_m$ must be introduced to eliminate the
logarithmic divergence of the integral. For free particles
(or sufficiently far above threshold) one finds for example
\begin{equation}\label{cut_dyn}
\Phi(E) = \sqrt{\frac{E}{m_t}} +
\frac{2 \Gamma_t}{\sqrt{m_t E}\pi}\ln \rmp_m/m_t
\end{equation}
This logarithmic divergence is a consequence of the fact
that the nonrelativistic approximation is used outside its
range of validity. On may either choose a cutoff of order
$m_t$ or replace the nonrelativistic phase space element
$\rmp\,d\rmp/m_t$ by $\rmp\,d\rmp/\sqrt{m_t^2 + \rmp^2}$. In practical
applications a cutoff will be introduced by the experimental
procedure used to define $t\bar t$-events.

\pagebreak[4]
\subsection{Polarization\label{pol_subs}}
\begin{figure}[h]
\begin{flushleft}
\leavevmode
\epsfxsize=8cm
\epsffile[100 370 500 520]{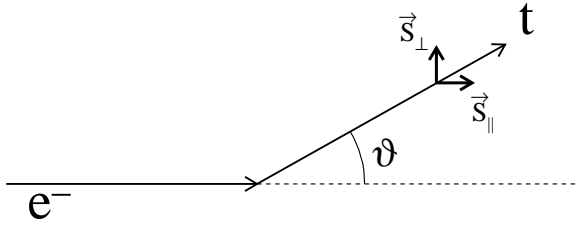}\\[-1.5cm]
\hfill
\parbox{6.cm}{\small
\caption[]{\label{dyn.ps}\sloppy Definition of the spin directions.
	The normal component ${\bf s}_{\rm N}$ points out of the plane.}}
\end{flushleft}
\end{figure}

To describe top quark production in the threshold region it
is convenient to align the reference system with the beam
direction (Fig.\ref{dyn.ps}) and to define
\begin{eqnarray}
{\bf s}_{\|} &=& {\bf n}_{e^-} \nonumber\\
{\bf s}_{\rm N} &=& {{\bf n}_{e^-} \times {\bf n}_t \over
	|{\bf n}_{e^-} \times {\bf n}_t|} 		\label{basis}\\
{\bf s}_\bot &=& {\bf s}_{\rm N} \times {\bf s}_{\|} .  \nonumber
\end{eqnarray}
%
In the limit of small $\beta$ the quark spin is essentially
aligned with the beam direction apart from small
corrections
proportional to $\beta$, which depend on the production
angle. A system of reference with ${\bf s}_\|$ defined with respect
to the top quark momentum\cite{krz}
is convenient in the high energy limit but evidently becomes
less convenient close to threshold.
\par\noindent
Including the QCD potential one obtains for the three components
of the polarization
\begin{eqnarray}
{\cal P}_\|(\bfp,E,\chi) &=& C_\|^0(\chi)
+ C_\|^1(\chi)\, \varphi_{_R}(\rmp,E)\,\cos\vartheta\,
 \label{thr_long}\\
{\cal P}_\bot(\bfp,E,\chi) &=& C_\bot(\chi)\,
\varphi_{_R}(\rmp,E)\,
\sin\vartheta\,
\label{thr_perp}\\
{\cal P}_{\rm N}(\bfp,E,\chi) &=& C_{\rm N}(\chi)
\varphi_{_I}(\rmp,E)
\sin\vartheta\,
	\label{thr_norm} ,
\end{eqnarray}\\
\parbox{75.ex}{
\begin{eqnarray*}
& &\hspace{5.ex}C_\|^0 (\chi) =
-{a_2 + \chi a_1 \over a_1 + \chi a_2} ,\hspace{6.9ex}
  C_\|^1 (\chi) = \left( 1-\chi^2 \right) {a_2 a_3 - a_1 a_4   \over
	\left(a_1 + \chi a_2 \right)^2} ,\\
& &\hspace{5.ex}C_\bot(\chi)  = -{1\over 2} \,
{a_4 + \chi a_3 \over a_1 + \chi a_2} ,
    \qquad C_{\rm N}(\chi) =-{1 \over 2}\, {a_3
    + \chi a_4 \over a_1 + \chi a_2}\, =\, - C_{\rm FB}(\chi) ,
\end{eqnarray*}}
\hfill
\parbox{5.ex}{
\begin{eqnarray} \label{coefs} \end{eqnarray} }
with $\varphi_{_I}(\rmp,E) = \Im\,\varphi$,  and $\varphi(\rmp,E)$
is defined in (\ref{phi}).
The momentum integrated quantities
are obtained by the replacement $\varphi(\rmp,E) \to \Phi(E)$. The case
of non-interacting stable quarks is recovered by the replacement
$\Phi\to\beta$, an obvious consequence of (\ref{cap_phi}).
\par
\begin{figure}[ht]
 \begin{center}
  \leavevmode
  \epsfxsize=14cm
  \epsffile[40 290 535 525]{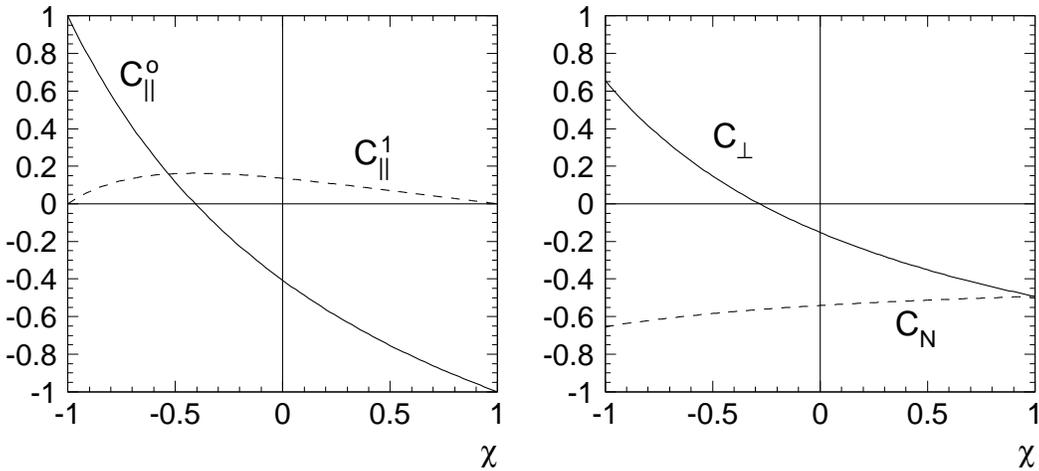}
  \caption[]{\label{pol_coefs.ps}\sloppy
	The coefficients (\ref{coefs}) for $\sqrt{s}/2 = m_t=180$ GeV.}
 \end{center}
\end{figure}

\noindent
Let us emphasize the
main qualitative features of the result.
\begin{itemize}
\item
Top quarks in the threshold region are highly polarized.
Even for unpolarized
beams the longitudinal polarization amounts to about $-0.41$
and reaches $\pm1$
for fully polarized electron beams. This later feature is of
purely kinematical
origin and independent of the structure of top quark couplings.
Precision
studies of polarized top decays are therefore feasible.
\item
Corrections to this idealized picture arise from the small admixture of
$P$-waves. The transverse and the normal components of
the polarization are of order 10\%. The angular dependent part
of the parallel polarization is even more suppressed.
Moreover, as a consequence of the
angular dependence its contribution
vanishes upon angular integration.
\item
The QCD dynamics is solely contained in
the functions $\varphi$ or $\Phi$
which is the same for the angular distribution and the various
components of the polarization. However, this ``universality''
is affected by the rescattering corrections\cite{HJKP}.
These functions which evidently depend on QCD dynamics can thus be
studied in a variety of ways.
\item
The relative importance of $P$-waves increases with energy,
$\Phi\sim\sqrt{E/m_t}$.
This is expected from the close analogy between $\Re\Phi$
and $\beta$.
\end{itemize}
The $C_i$ are displayed
in Fig.\ref{pol_coefs.ps}
as functions of the polarization $\chi$.
For the weak mixing angle a value
$\sin^2\!\theta_W= 0.2317$ is adopted,
for the top mass $m_t=180$ GeV.  As
discussed before, $C_\|^0$ assumes its maximal value $\pm 1$ for $\chi=\mp 1$
and the coefficient $C_\|^1$ is small throughout. The coefficient $C_\bot$
varies between $+0.7$ and $-0.5$ whereas $C_{\rm N}$ is typically around
$-0.5$.  The dynamical factors $\Phi$ are around $0.1$ or
larger, such that the $P$-wave induced effects should be observable
experimentally.
\par\noindent
The normal component of the polarization which is proportional to
$\varphi_{_I}$ has been predicted for stable quarks in the framework of
perturbative QCD \cite{dev,krz}. In the threshold region the phase can be
traced to the $t\bar t$ rescattering by the QCD potential. For a pure
Coulomb potential $V=-4\alpha_s/3r$ and stable quarks the
nonrelativistic problem can be solved analytically \cite{FKotsky} and
one finds
\begin{eqnarray}
\varphi_{_I}(\rmp,E) &\rightarrow&
{2\over 3}\alpha_s{1-4\alpha_s/3\pi\over
	1-8\alpha_s/3\pi} \label{phi_gt0}\\
\Phi_{_I}(\rmp,E) &\rightarrow&
{2\over 3}\alpha_s{1-4\alpha_s/3\pi\over
	1-8\alpha_s/3\pi} . \label{cap_phi_gt0}
\end{eqnarray}
The component of the polarization normal to
the production plane is thus
approximately independent of $E$ and essentially measures the strong
coupling constant. In fact one can argue that this is a unique way to
get a handle on the scattering of heavy quarks through the QCD
potential.

\section{Acknowledgments}

M.J. would like to thank the organizers of the Workshop for
their extraordinary hospitality and successful efforts to
create stimulating atmosphere in Morioka. He is particularly
grateful to Professors Hitoshi Murayama and Yukinari Sumino,
and to Dr.~Takehiko Asaka for helpful scientific discussions
and for fascinating excursions.

\def\app#1#2#3{{\elevenit Act.~Phys.~Pol.~}{\elevenbf B #1} (#2) #3}
\def\apa#1#2#3{{\elevenit Act.~Phys.~Austr.~}{\elevenbf#1} (#2) #3}
\def\fortp#1#2#3{{\elevenit Fortschr.~Phys.~}{\elevenbf#1} (#2) #3}
\def\npb#1#2#3{{\elevenit Nucl.~Phys.~}{\elevenbf B #1} (#2) #3}
\def\plb#1#2#3{{\elevenit Phys.~Lett.~}{\elevenbf B #1} (#2) #3}
\def\prd#1#2#3{{\elevenit Phys.~Rev.~}{\elevenbf D #1} (#2) #3}
\def\pR#1#2#3{{\elevenit Phys.~Rev.~}{\elevenbf #1} (#2) #3}
\def\prl#1#2#3{{\elevenit Phys.~Rev.~Lett.~}{\elevenbf #1} (#2) #3}
\def\prc#1#2#3{{\elevenit Phys.~Reports }{\elevenbf #1} (#2) #3}
\def\cpc#1#2#3{{\elevenit Comp.~Phys.~Commun.~}{\elevenbf #1} (#2) #3}
\def\nim#1#2#3{{\elevenit Nucl.~Inst.~Meth.~}{\elevenbf #1} (#2) #3}
\def\pr#1#2#3{{\elevenit Phys.~Reports }{\elevenbf #1} (#2) #3}
\def\sovnp#1#2#3{{\elevenit Sov.~J.~Nucl.~Phys.~}{\elevenbf #1} (#2) #3}
\def\yadfiz#1#2#3{{\elevenit Yad.~Fiz.~}{\elevenbf #1} (#2) #3}
\def\jetp#1#2#3{{\elevenit JETP~Lett.~}{\elevenbf #1} (#2) #3}
\def\zpc#1#2#3{{\elevenit Z.~Phys.~}{\elevenbf C #1} (#2) #3}
\def\ptp#1#2#3{{\elevenit Prog.~Theor.~Phys.~}{\elevenbf #1} (#2) #3}
\def\nca#1#2#3{{\elevenit Nouvo~Cim.~}{\elevenbf #1A} (#2) #3}
\section{References}
\vspace{-0.5cm}
\vglue 0.4cm

\end{document}